\documentclass{iaus}
\usepackage{graphicx}

\usepackage{natbib}

\newcommand\aap{\textit{A\&A}}
\newcommand\aaps{\textit{A\&AS}}
\newcommand\mnras{\textit{MNRAS}}
\newcommand\apj{\textit{ApJ}}
\newcommand\apjl{\textit{ApJ}}
\newcommand\apjs{\textit{ApJ}}
\newcommand\aj{\textit{AJ}}

\newcommand\ulsr{U_{\rm LSR}}
\newcommand\vlsr{V_{\rm LSR}}
\newcommand\wlsr{W_{\rm LSR}}

\newcommand\teff{T_{\rm eff}}

\title[Characterisation of the Galactic thick disk] 
{Characterisation of the Galactic thick disk}

\author[Thomas Bensby]   
{Thomas Bensby$^1$}%

\affiliation{$^1$Lund Observatory, Box 43, SE-221\,00 Lund, Sweden}

\pubyear{2014}
\volume{298}  
\pagerange{000--000}
\jname{Setting the scene for Gaia and LAMOST}
\editors{S. Feltzing, G. Zhao, N.\,A. Walton,\& P.\,A. Whitelock , eds.}
\begin{document}

\maketitle

\begin{abstract}
Thick disks appear to be common in external large spiral galaxies 
and our own Milky Way also hosts one. The existence of a thick disk 
is possibly directly linked to the formation history of the host 
galaxy and if its properties is known it can constrain models of 
galaxy formation and help us to better understand galaxy formation 
and evolution. 
This brief review attempts to highlight some of the characteristics 
of the Galactic thick disk and how it relates to other stellar 
populations such as the thin disk and the Galactic bulge. Focus has 
been put on results from high-resolution spectroscopic data obtained 
during the last 10 to 15 years.
\end{abstract}

\firstsection 

\section{Introduction}

More than three decades ago, \cite{tsikoudi1979} and \cite{burstein1979}
found the first observational evidence for second, and thicker,
disk components in external edge-on galaxies. A few years later 
\cite{gilmore1983} found the first 
evidence for a second stellar component in the Milky Way disk by 
studying the 
stellar density as a function of distance from the Galactic plane 
towards the Galactic North pole. Since then the general consensus has been
that also the Milky Way harbours a dual disk stellar system, a thin disk
and a thick disk. Recent studies actually show that many, if not all, 
edge-on spiral galaxies appears to host dual disk systems 
\citep{yoachim2006,comeron2011}. As the Milky Way currently is the only 
galaxy that can be studied in great detail with high-resolution 
spectrographs, and may serve as a ``benchmark galaxy'' for extra-galactic 
studies, it is utterly important, also in the context of galaxy formation 
and as tests of models of galaxy formation, to establish the properties 
of the different Milky Way stellar populations.

Since the discovery of the Galactic thick disk many observational 
studies have aimed at characterising the stellar disk 
\citep[e.g.,][]{edvardsson1993,feltzing1998} and later especially
targeting the thick disk, and its differences relative to the thin disk
\citep[e.g.,][]{fuhrmann1998,fuhrmann2000unpubl,fuhrmann2004,fuhrmann2008,fuhrmann2011,
prochaska2000,gratton2000,chen2000,mashonkina2001,tautvaisiene2001,
bensby2003,bensby2004,bensby2005,bensby2007letter2,
soubiran2003,reddy2003,reddy2006,adibekyan2012}.
The observational evidence presented by these studies have so far
pointed to two disk populations
with  different chemical and age properties indicating that the 
Galactic thin and
thick disk have different origins and have experienced
different chemical histories.

\begin{figure}
\resizebox{\hsize}{!}{
  \includegraphics[bb=50 -10 540 820, clip, angle=-90]{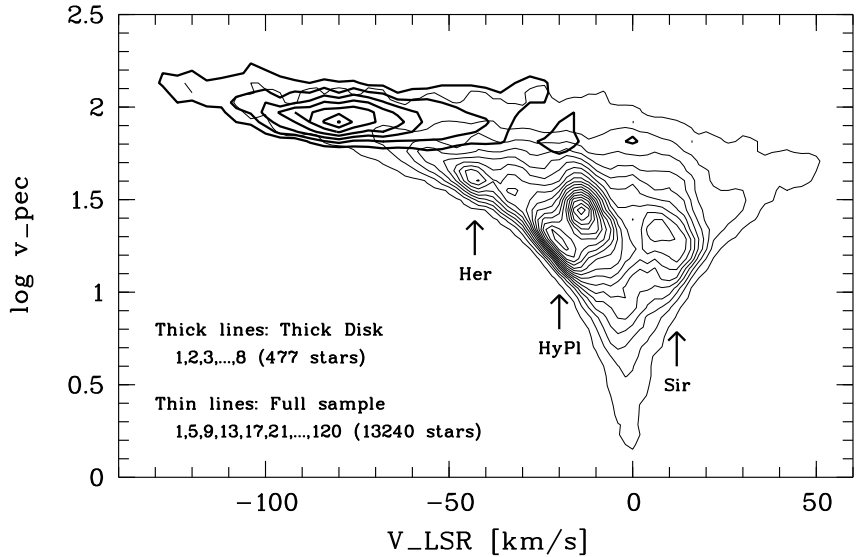}}
\caption{
        Velocity distribution of 16\,000 dwarf stars in the
        GCS
        (where $v_{\rm pec}\equiv(\ulsr^2 + \vlsr^2 + \wlsr^2)^{1/2}$).
        The Hercules stream (Her), the Hyades-Pleiades cluster (HyPl),
        and the Sirius group (Si) have been marked. Distances between the
        isodensity curves are also given in the plot. {\sl Figure from
        \cite{bensby2007letter}}.
        }
  \label{fig:contour}
\end{figure}

However, albeit more than two decades of observational effort we are still 
lacking much information about the complex abundance structure of the 
Galactic stellar disk. For instance,
the Geneva-Copenhagen Survey (hereafter GCS) by \cite{nordstrom2004}
contains approximately 16\,000 dwarf stars in the Solar 
neighbourhood, all of which have full three-dimensional kinematic 
information available, as well as ages and metallicities estimated from
Str\"omgren $uvby \beta$ photometry. It is evident from the GCS
that there is a lot of kinematic sub-structure in the solar neighbourhood
(see Fig.~\ref{fig:contour}). While the thin and thick disks are 
the dominating populations there are many smaller structures 
superimposed on the underlying disk distribution. Most of these are 
young streams and moving groups that belong to the thin disk, but 
there are also features such as the Hercules stream that share some 
of the thick disk kinematic properties \citep[e.g.][]{famaey2005}.
It has later been shown that the Hercules stream most likely is
a dynamical feature in velocity space caused by the Galactic bar, 
and that it consists of a mixture of thin and thick disk stars 
\citep[e.g.,][]{bensby2007letter}.

\begin{figure}
\resizebox{\hsize}{!}{
  \includegraphics[bb=-32 154 642 520, clip]{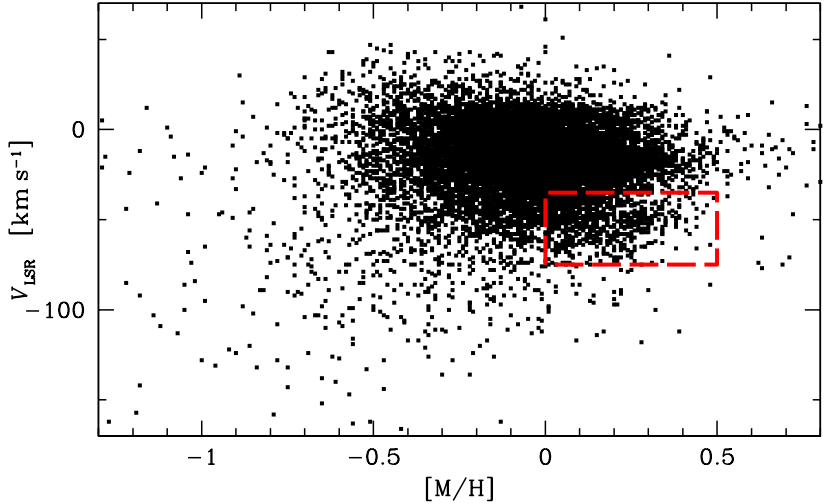}}
\caption{$\vlsr$ versus [M/H] for the 16\,000 stars in the GCS.
The metallicities are taken from the new calibration by \cite{casagrande2011}.
The dashed rectangle highlight the area where there might be metal-rich
thick disk stars.
                }
  \label{fig:vfeh}
\end{figure}

Looking at the GCS data it is also evident that
stars with typical thick-disk kinematics can be found at high 
metallicities, even well above solar (see Fig.~\ref{fig:vfeh}).
The question is where the high-metallicity limit of the thick
disk is and if these stars are true thick disk stars? 
It is also not clear to how low metallicities the thin disk reach, 
and if there is a gap in the abundance trends between the thin and 
thick disks. 

I will here highlight some of the characteristics of the Galactic thick
disk with focus on results from high-resolution spectroscopic
studies during the last 10 to 15 years. This is no
exhaustive and full summary, and  there are several other aspects 
of the thick disk that needs further attention.

\section{Properties of the Galactic thick disk}

\subsection{General properties}

\begin{figure}
\resizebox{\hsize}{!}{
  \includegraphics[bb=-15 154 652 465, clip]{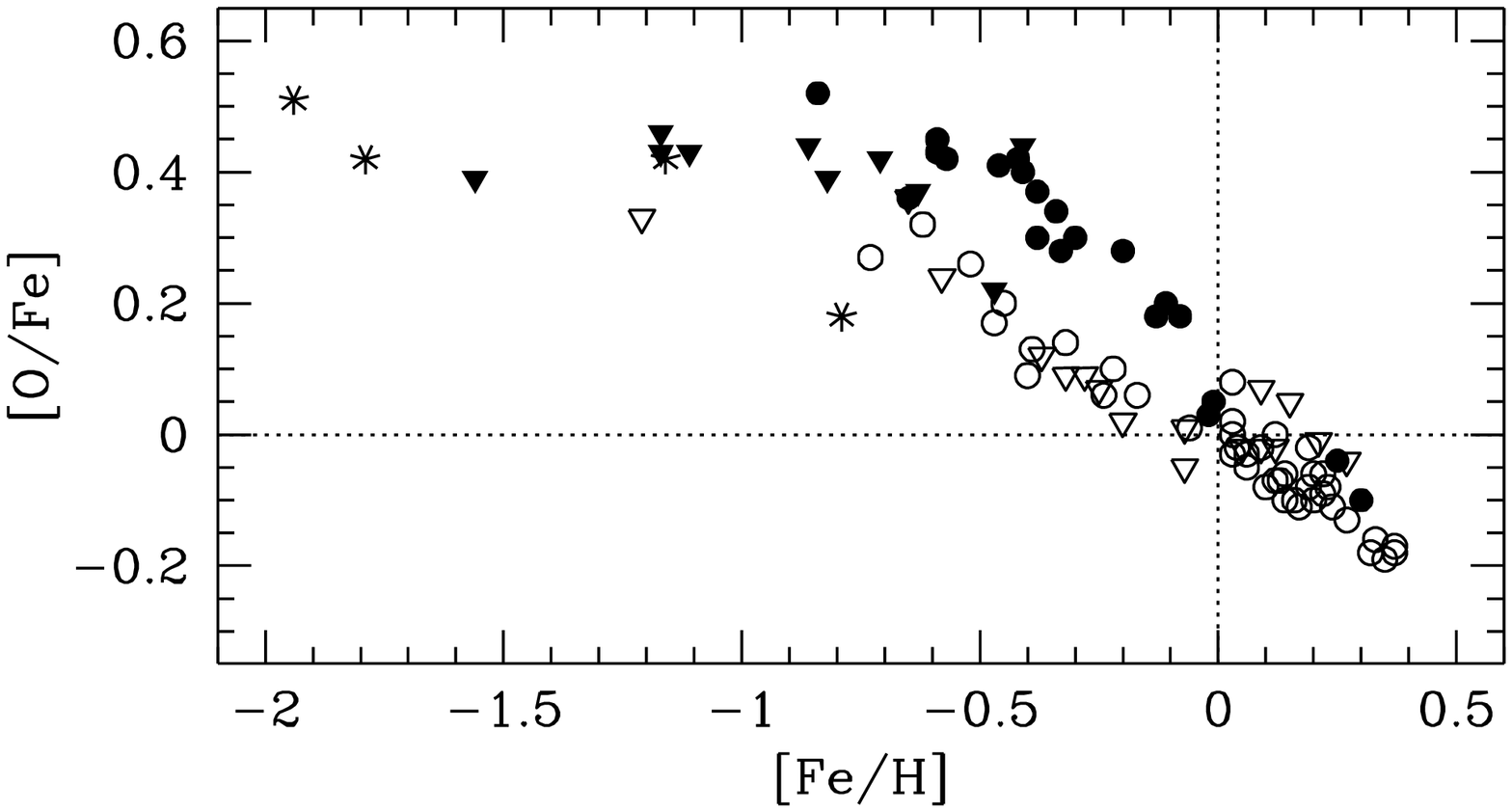}}
\caption{[O/Fe] versus [Fe/H] for thin disk stars (open symbols)
and thick disk stars (filled symbols). Data from \cite{bensby2004}
and \cite{nissen2002}. {\sl Figure from \cite{bensby2004}}.
                }
  \label{fig:ofe}
\end{figure}

In summary, the general consensus regarding the thick disk is that it
has a metallicity distribution (MDF) that peaks around 
$\rm [Fe/H]\approx -0.6$ \citep[e.g.,][]{gilmore1995,carollo2010}, 
while the thin disk MDF peaks at $\rm [Fe/H]\approx 0$ 
\citep[e.g.,][]{casagrande2011}. At a given metallicity the
thick disk stars show higher enhancement in the $\alpha$-element 
abundances than the thin disk stars \citep[e.g.,][]{prochaska2000,reddy2003,soubiran2003,bensby2003,bensby2004,bensby2005,reddy2006,bensby2007letter2,adibekyan2012}. For an $\alpha$-element, such
as O, Mg, Si, Ca, or Ti, the [$\alpha$/Fe] abundance ratio for the
thick disk shows a
flat plateau from $\rm [Fe/H]\approx -1$ to $-0.4$, hereafter
it starts to decline toward solar values. The plateau is caused
by the rapid chemical enrichment from massive stars, while the down-turn,
or ``knee'' is attributed to the onset of enrichment from SN\,Ia.
The appearance of the thin disk [$\alpha$/Fe] abundance trends are quite
different, starting out a lower enhancement, and then showing a constant
slow decline toward super-solar metallicities. An example 
of the thin and thick disk abundance trends is shown for oxygen
in Fig.~\ref{fig:ofe}. 

Furthermore, the stars of the thick disk are generally old with ages 
between 8 to 12\,Gyr, in contrast to the stars of the thin disk that 
usually are young \citep[e.g.,][]{bensby2003,schuster2006,haywood2006}.
It may also contain an age-metallicity relation 
\citep[e.g.,][]{bensby_amr} which the thin disk does not have.

The existence, or non-existence, of radial and 
vertical abundance gradients in the thick disk is difficult to establish
due to the problems of selecting clean thin and thick disk stellar 
samples. However, using the SDSS SEGUE dwarf sample \cite{cheng2012_1} 
see a flat radial gradient for stars located 1 to 1.5\,kpc from the 
plane that most likely can be associated with the thick disk. 
Also, from 22\,000 stars in SDSS DR3 \cite{allendeprieto2006}
finds a flat radial metallicity gradient for the thick disk between 
galactocentric radii 4 to 15\,kpc, and furthermore that
no vertical abundance gradient is apparent for the thick disk.

\subsection{A nearby volume-complete sample}

Klaus Fuhrmann has in a series of papers \citep{fuhrmann1998,fuhrmann2000unpubl,fuhrmann2004,fuhrmann2008,fuhrmann2011} 
analysed what is quoted as a volume-complete sample of all nearby 
($d\leq 25$\,pc) mid-F-type to early K-type stars down to 
$M_{\rm V} = 6.0$ and north of a declination of $\delta = -15^\circ$. 
This analysis is particularly valuable thanks to its internal 
consistency and homogeneity. In these papers Fuhrmann finds that 
there are essentially two types of stars in his sample: stars with 
high abundances of Mg relative to Fe, and stars with low abundances 
of Mg relative to Fe. The two subsets overlap significantly in [Fe/H]. 
The stars with higher [Mg/Fe] abundance ratios are also the oldest 
stars. Both sets of stars, but in particular the low [Mg/Fe] stars, 
show a very tight abundance trend for [Mg/Fe] versus [Fe/H]. 
Fuhrmann identifies these two sub-samples with the thick and thin 
disks, respectively. The identification is based on an interpretation 
of a combination of the age, kinematic and abundance data. 

In addition a few stars are found to show [Mg/Fe] ratios that 
are in-between those of the two major sub-samples. These are given 
the status of ``transition'' stars and are considered to be neither
thin nor thick disk members.

The most important part of these studies by Fuhrmann
is that they show that if 
a reasonably volume-complete sample is constructed for stars in the 
vicinity of the Sun, then the elemental abundances (and ages) slit 
these stars into two major groups, one with enhanced [Mg/Fe] ratios, 
and one with almost solar values. However, the Fuhrmann sample
only contains slightly over 20 stars out of almost 400 stars that have 
age and abundance patterns that make them potential thick disk stars, 
and they only reach metallicities slightly above $\rm [Fe/H]\approx -0.3$. 
What about the kinematically hot and metal-rich stars that are present 
in the GCS? These are clearly not present in the very nearby sample, and 
it is evident that in order to study the thick disk at its 
extremes we need to probe a larger volume than Fuhrmann's.

\subsection{Kinematically selected samples}

\begin{figure}
\resizebox{\hsize}{!}{
\includegraphics[bb=-72 144 682 450,clip]{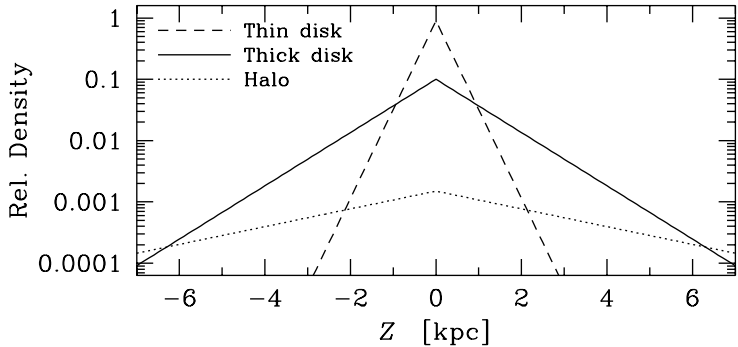}}
\caption{Relative stellar density as a function of distance from the
Galactic plane (assuming scale-heights of 300\,pc and 1000\,pc,
and normalisations of 90\,\% and 10\,\%, for the thin and thick
disks, respectively). {\sl Figure from \cite{bensby2004phdt}.}
\label{fig:stellardensity}}
\end{figure}

In the Solar neighbourhood, in the Galactic plane, more than
90\,\% of the stars are believed to be thin-disk stars,
and less than 10\,\% thick-disk stars.
However, as the thin disk has a much shorter scale-height than
the thick disk, the relative stellar density between the
two populations will vary with distance from the plane.
Assuming scale-heights of 300\,pc and 1000\,pc,
and normalisations of 90\,\% and 10\,\%, for the thin and thick
disks, respectively, Fig.~\ref{fig:stellardensity} shows how the
stellar density varies as a function of distance from the
Galactic plane. One sees that at a distance of 1\,kpc from the
plane the thin and thick disk stellar densities are about the same,
and that at distances greater than 2\,kpc the thick disk
clearly dominates over the thin disk. Hence one needs to go to distances 
of at least $\sim 2$\,kpc from the plane in order to get a relatively 
clean thick disk sample (the stellar halo starts to dominate
much farther from the plane). However, at these distances
F and G dwarf stars have magnitudes around 16 to 17 making
it very time consuming to get high-resolution spectra with high
signal-to-noise ratios. They are simply too faint even for today's
large 8-10\,m telescopes to observe in large quantities. 
 
What has instead been done the past 10 to 15 years is to 
rely on kinematic criteria to select candidate thick disk stars 
from nearby  stellar samples (typically within 100-200\,pc from the 
Sun, i.e., the Hipparcos sphere) \citep[see, e.g.,][]{bensby2003}. 
In order to calculate probabilities
one needs an estimation of the overall velocity distribution (and
assuming Gaussian distributions) of the
stellar population, how much its lags behind the local
standard of rest (LSR), and its relative stellar density
in the solar neighbourhood. The best values for the velocity distributions,
rotational lags, and normalisations, are often difficult to estimate,
but some commonly used values are listed
in Table~\ref{tab:kinematic}.  

\begin{table}[t]
\centering
\caption{
\label{tab:kinematic}
        Characteristics for stellar populations 
        in the Solar neighbourhood. Columns (2)-(4) give the velocity 
        dispersions for the different populations
		in col.~(1); cols.~(5)-(6) give the the asymmetric drifts 
		(in $U$ and $V$) relative to the LSR; and col.~(7) gives the 
        normalisation fractions for each population in the Solar neighbourhood
        (in the Galactic plane).
        Values are taken from
        \cite{bensby2005,bensby2007letter,famaey2005} for the thin disk, thick disk, 
        the stellar halo, and the Hercules stream.
        }
\begin{tabular}{lrrrrrl}
\noalign{\smallskip}\hline \hline\noalign{\smallskip}
        & $\sigma_{\rm U}$
        & $\sigma_{\rm V}$
        & $\sigma_{\rm W}$
        & $U_{\rm asym}$
        & $V_{\rm asym}$ 
        & $X$  \\
\noalign{\smallskip}
        & \multicolumn{5}{c}{-----------~~[km\,s$^{-1}$]~~-----------}     
        & \\
\noalign{\smallskip}
\hline\noalign{\smallskip}
   Thin disk   &  35  & 20 & 16 &    0  &  $-15$ &  0.85  \\
   Thick disk  &  67  & 38 & 35 &    0  &  $-46$ &  0.09  \\
   Halo        & 160  & 90 & 90 &    0  & $-220$ &  0.0015 \\
   Hercules    &  26  &  9 & 17 & $-40$ &  $-50$ &  0.06    \\
\hline
\end{tabular}
\end{table}

Applying these criteria to a catalogue such as for example the 
Geneva-Copenhagen survey (GCS) \citep{nordstrom2004}, that
contains roughly 16\,000 nearby dwarf stars with full three-dimensional
kinematical information allows you to
calculate probabilities for individual stars of belonging
different populations \citep[see, e.g.,][]{bensby2003}.
As the GCS also contains metallicities and ages based
on Str\"omgren $uvby$ photometry for a majority of the 16\,000 stars, give
the opportunity to probe the thin and thick disks at their
extreme metallicities. 

Studies that have utilised kinematical criteria to investigate
the properties of the thin and thick disks are, e.g., 
\cite{prochaska2000,reddy2003,soubiran2003,bensby2003,bensby2004,bensby2005,reddy2006,bensby2007letter2,adibekyan2012}, wherein nice figures 
can be found 
that show the elemental abundance trends of the thin and thick
disks. In summary, the studies above generally find that
the thin and thick disks have different abundance trends. 

A shortcoming of the kinematical method to classify stars into
different populations is that it
will introduce bias, or mixing, between the selected thin 
and thick disk samples. Stars from the high-velocity tail of the
thin disk will be classified as thick disk stars and stars from the
low-velocity tail of the thick disk as thin disk stars. 
In \cite{bensby2013disk} this is investigated and it is
shown that stellar age might be a better discriminator between
the thin and thick disks. By selecting one sample of stars that are 
older than $\sim8$\,Gyr and one where the stars are younger than
$\sim 8$\,Gyr, the abundance trends become apparently cleaner
with less mixing between the two populations
(see, e.g., Fig.~2 in \citealt{bensby2010rio}). The problems with using
stellar ages is that good ages are only
possible to determine for stars near or around the main sequence turn-off and on
the sub-giant branch. For red giant samples or more distant samples kinematical
criteria might be the only solution. One should then bear in mind that
kinematic mixing could be significant. 

\subsection{The dichotomy of the Galactic disk}

A question that has recently surfaced and been debated is 
if the Milky Way disk really has two distinct stellar populations. 
Based on the SDSS SEGUE G and K dwarf stellar sample
\citep{abazajian2009,yanni2009}, \cite{bovy2012}
instead argues that the abundance pattern
of the Galactic stellar disk can be represented by a continuous  
function of mono-abundance populations with increasing scale-heights,
and hence no {\it distinct} thick disk should be claimed.

\begin{figure}
\resizebox{\hsize}{!}{
\includegraphics[bb=-72 144 682 718,clip]{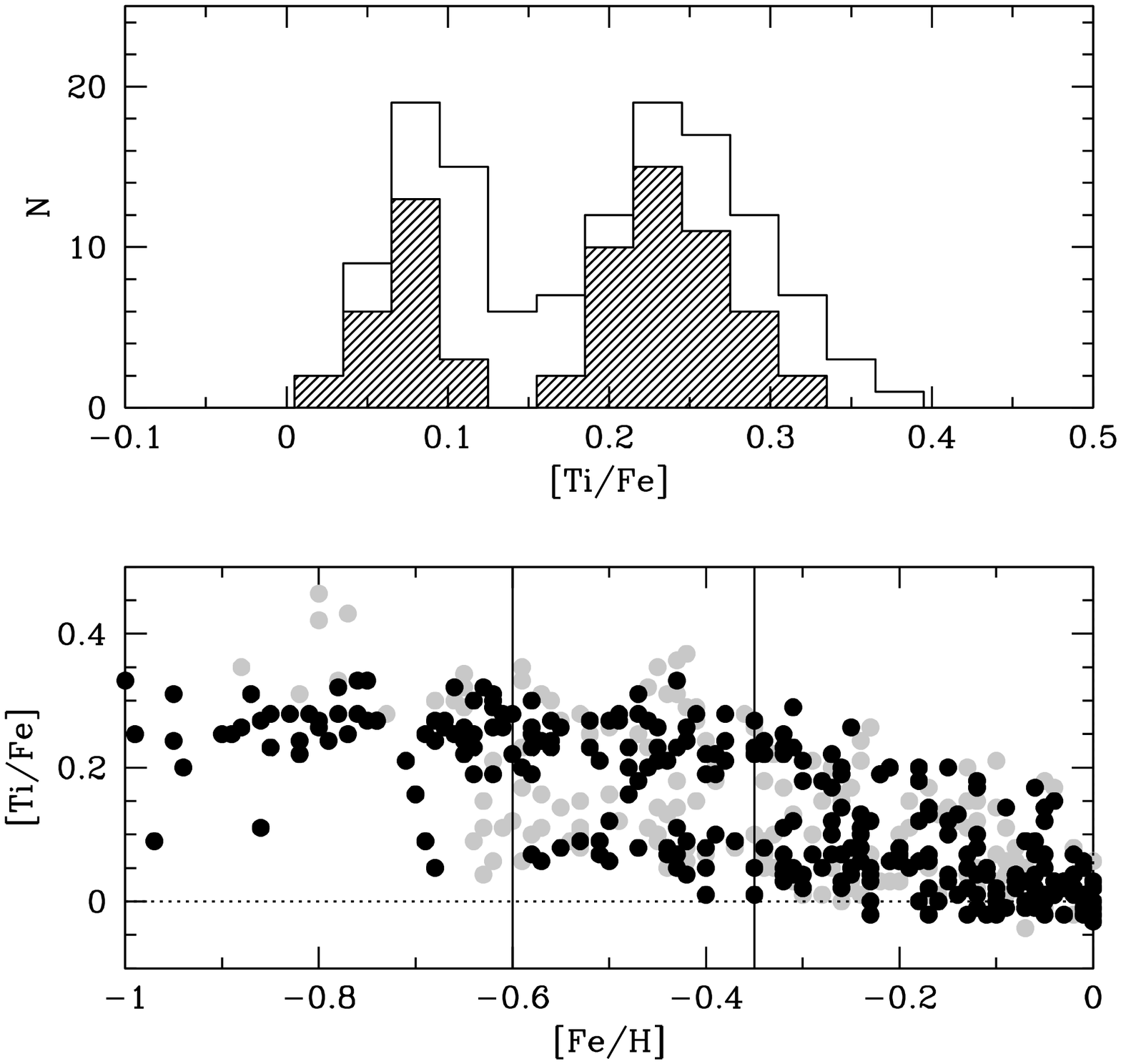}}
\caption{Bottom panel shows the [Ti/Fe]-[Fe/H] abundance
plot of the 700 nearby F and G dwarf stars from 
\cite{bensby2013disk} in the metallicity range $\rm -1<[Fe/H]<0$
for stars with $\teff$ between 5600-6100\,K marked by black circles,
and stars outside this range with grey circles. The top panel show
the [Ti/Fe] histogram for the stars with metallicities
between $\rm -0.6<[Fe/H]-0.35$. White histogram include all stars
in this metallicity range, while the dashed histogram is further constrained
to the stars in the 5600-6100\,K temperature range.
\label{fig:gap}}
\end{figure}

If truly distinct and well-separated 
abundance trends between the thin and thick disks could 
be confirmed, the claims by \cite{bovy2012} could be weakened. 
Using the data sample of 1111 stars by \cite{adibekyan2012}, clues of a 
two-phase formation history of the Milky Way disk is seen
by \cite{haywood2013}. Additionally, \cite{bensby2013disk}
see a void of stars in the $\alpha$-element abundance trends
in a sample of 700 kinematically selected F and G dwarf stars.
This void remains, even if stars with kinematic properties
in-between the two disks are included. It becomes even clearer and more
separated if the sample is constrained to a narrow range in temperature
between 5600 and 6100\,K (see Fig.~\ref{fig:gap}). The stellar
parameters and elemental abundances for stars outside
this temperature range are more susceptible to NLTE effects and 
uncertainties \citep[see][for discussion]{bensby2013disk}.
This shows that small signatures, such as the separation between
the thin and thick disk abundance trends, can be smeared out if 
sufficient care is not taken in the analysis and sample selection.

Further clues to a distinction between the thin and thick disks
is hinted in their age structures. For instance, most thick disk
stars have ages around 10\,Gyr, but as you go to higher metallicities
they become slightly younger, and the most metal-rich 
stars that can be associated with the thick disk at 
$\rm [Fe/H]\approx 0$ have 
ages of about 8\,Gyr \citep{bensby2007letter2}. The stars of the thin disk are, on the other 
hand, younger, with the oldest stars at $\rm [Fe/H]\approx -0.7$ being
around 7\,Gyr.

This time epoch where we see a separation between the two disks, 
around 8\,Gyr
ago, coincides with other observational evidence for mergers between the 
Milky Way and another (dwarf) galaxy. For instance, \cite{gilmore2002}
and \cite{wyse2006}
claims to have detected debris stars from a major merger $\sim$10\,Gyr
ago, and \cite{deason2013} finds that the density profile of the Milky Way
halo is broken, and that this break likely is associated with a an early 
(6-9 Gyr ago) and massive accretion event.

\section{The inner and outer disk}

\begin{figure*}
\centering
\resizebox{\hsize}{!}{
\includegraphics[bb= 0 160 445 620,clip]{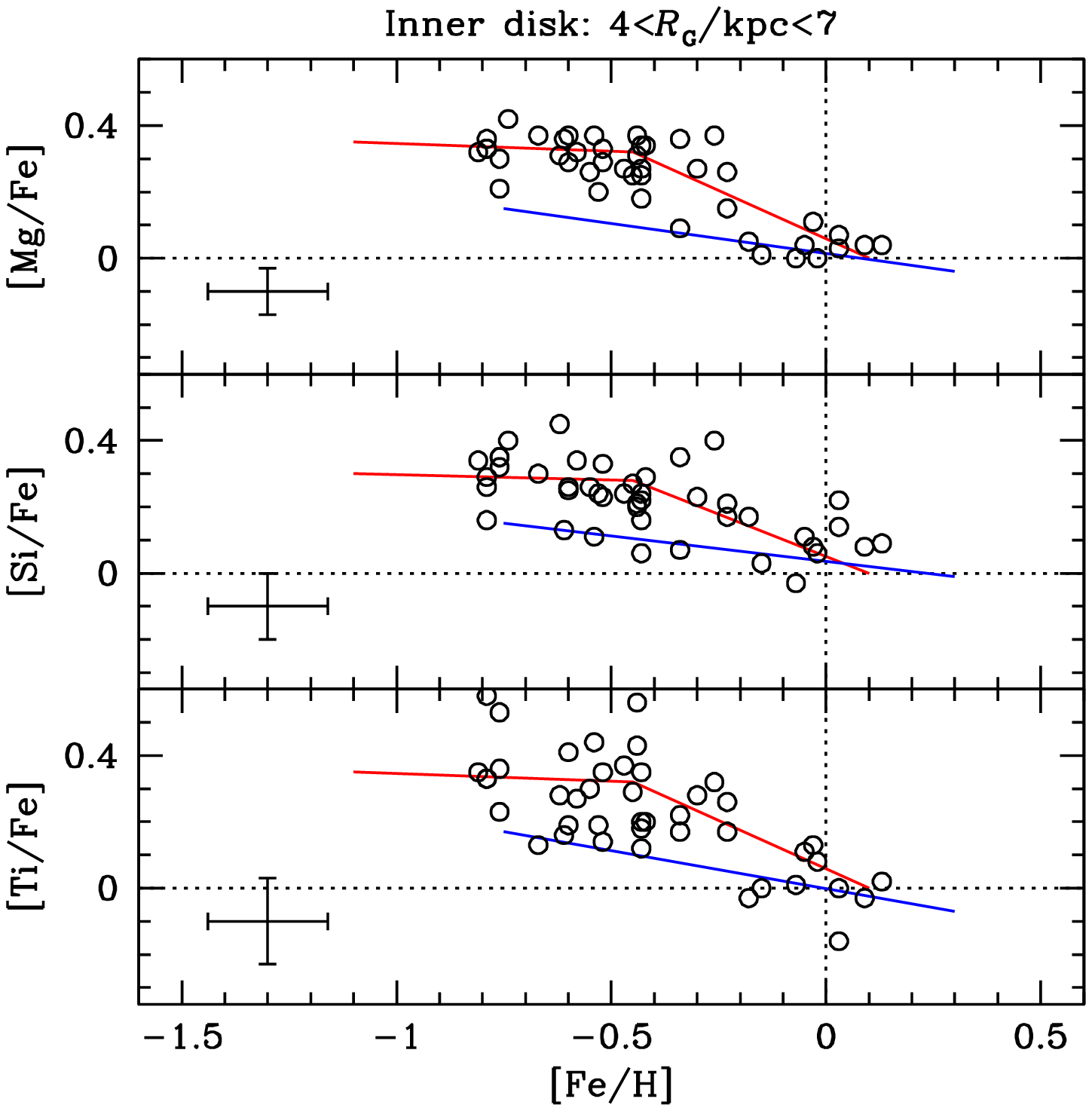}
\includegraphics[bb=75 160 445 620,clip]{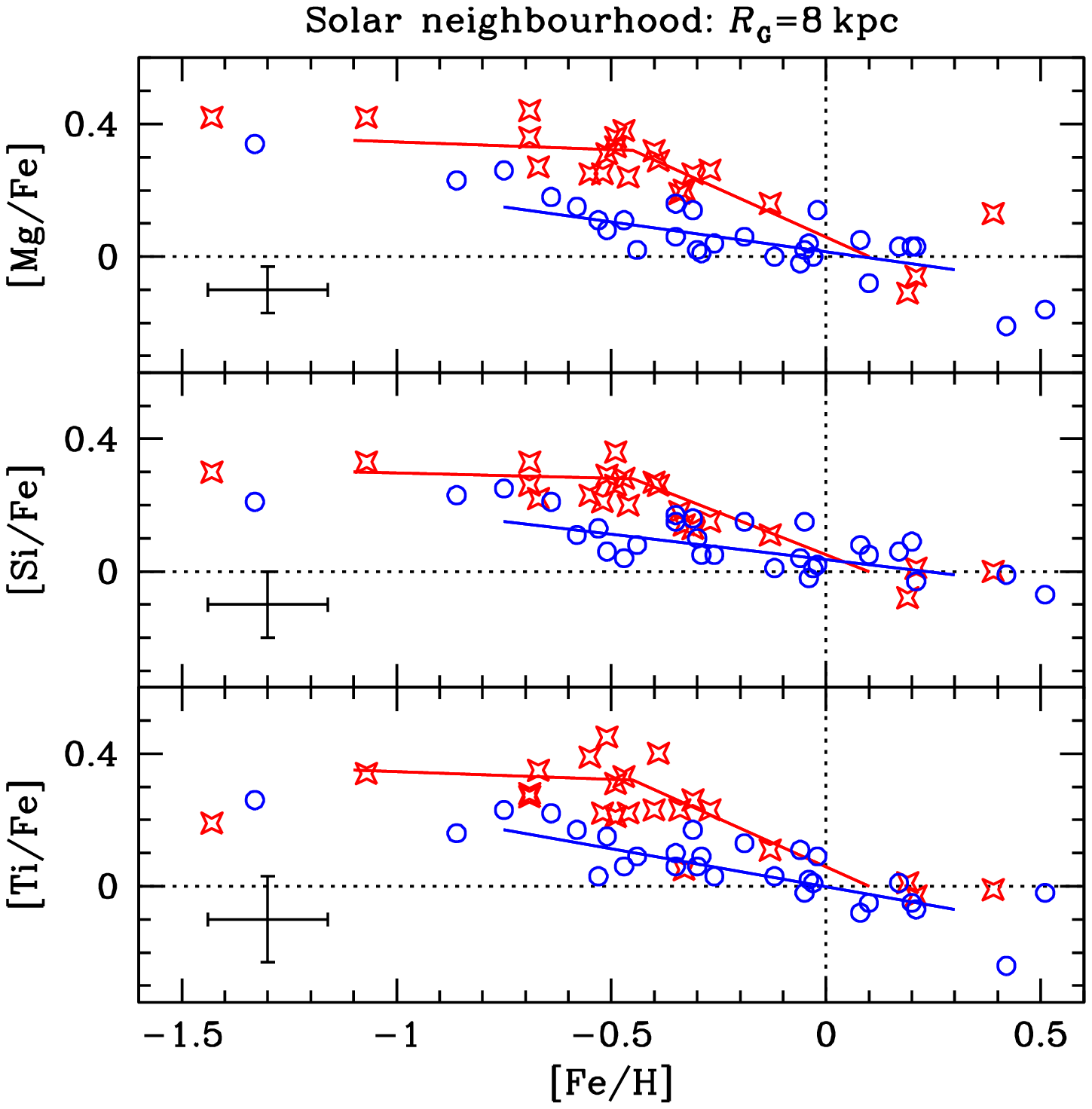}
\includegraphics[bb=75 160 465 620,clip]{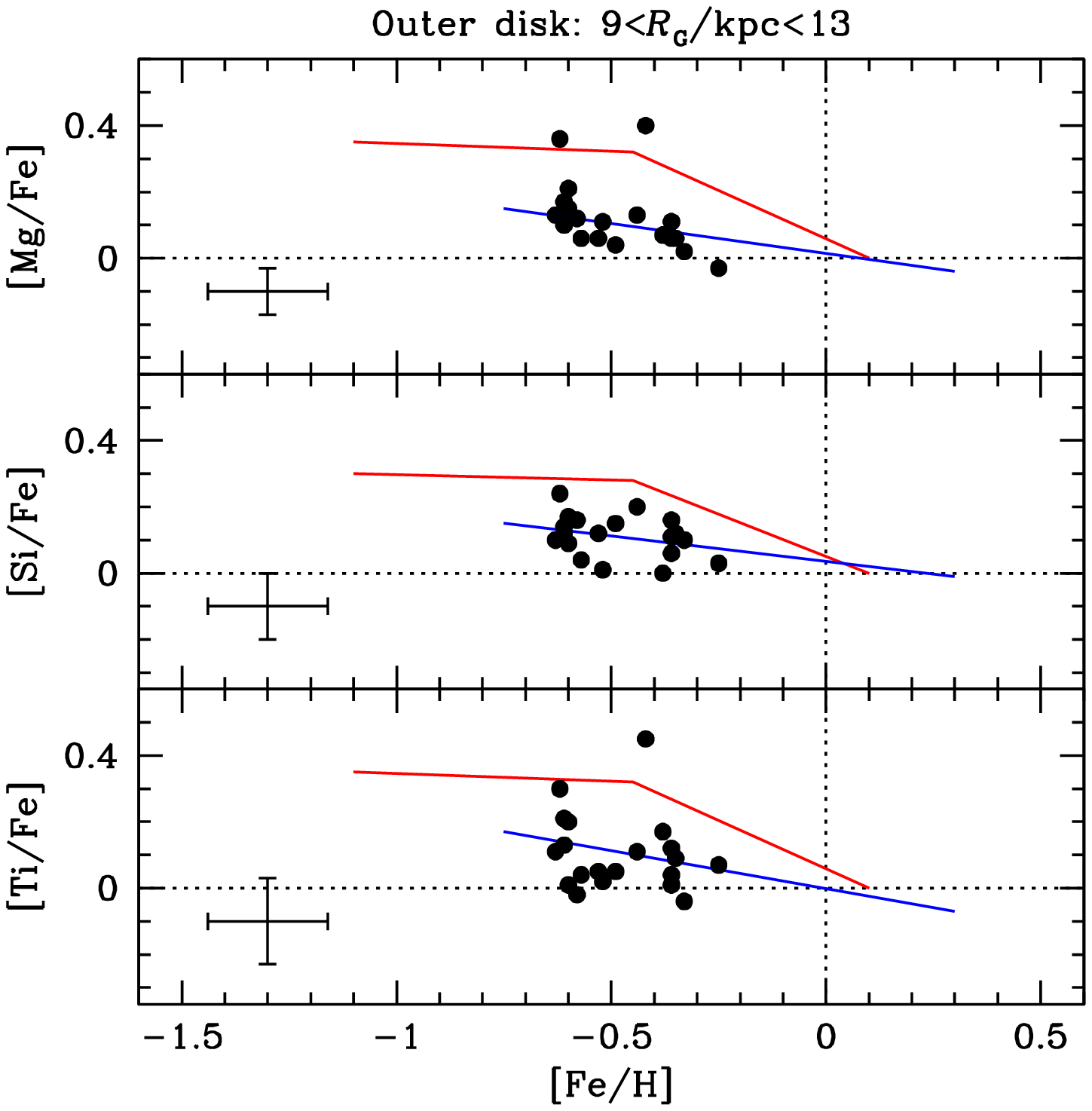}}
\caption{
Abundance trends for the $\alpha$-elements Mg, Si, and Ti
for 44 inner disk red giants (left panel) from 
\citep{bensby2010letter}, nearby thin and thick disk red giants  
(blue circles and red stars, respectively, middle panel) from 
\cite{alvesbrito2010}, and 20 outer disk red giants (right panel)
from \cite{bensby2011letter}. {\sl Figure from \cite{bensby2011letter}}. 
\label{fig:outerdisk} 
}
\end{figure*}

The thick disk and its relationship to the
thin disk has so far mainly been studied at the solar 
galactocentric radius. The existence of the thick disk
in the inner and outer 
regions of the Galactic disk is less established.
In a first study \cite{bensby2010letter} observed 44 red giants
in the inner disk at galactocentric radii between 4 to 7\,kpc
and at different heights from the Galactic plane. Those results showed
that the abundance trends seen in the solar neighbourhood are also 
seen in the inner disk region, pointing to that a dual disk structure
is also present in the inner disk (see left panel in 
Fig.~\ref{fig:outerdisk}). In a subsequent similar study of 
20 red giants in the outer disk \cite{bensby2011letter}, the
abundance structure appeared very different compared to the
inner disk results
(see right panel in Fig.~\ref{fig:outerdisk}). There were essentially
no traces of the abundance signatures that usually are 
associated with the thick disk and which is seen in the inner disk
and in the solar neighbourhood. The conclusion from \cite{bensby2011letter}
was that the scale-length of the thick disk is significantly
shorter than that of the thin disk, and hence the apparent
lack of thick disk stars in the outer disk. Evidence
for a short scale-length for the thick disk was later also seen in
the SDSS G dwarf sample by \cite{cheng2012_2}.

\section{Similarities to the Galactic bulge}

\label{sec:bulge}

\begin{figure}
\resizebox{\hsize}{!}{
\includegraphics[bb=-72 385 682 718,clip]{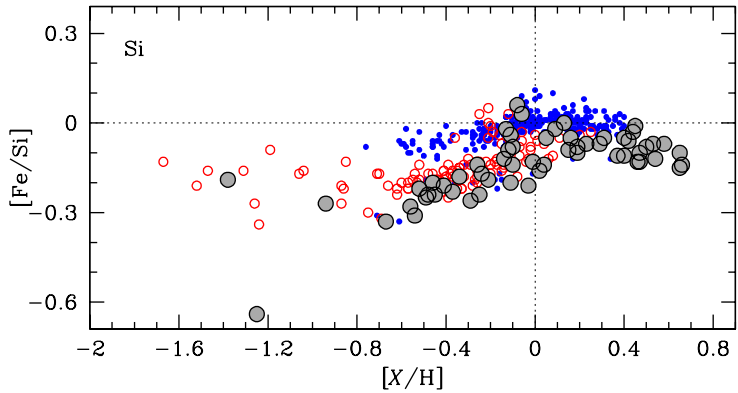}}
\caption{[Fe/Si] versus [Si/H] for 58 microlensed dwarf and subgiant
stars from \cite{bensby2013} (grey circles), 
nearby thick disk dwarf stars (red circles) and nearby thin disk 
dwarf stars (blue dots) from \cite{bensby2013disk}. {\sl Figure 
from \cite{bensby2013}}.
\label{fig:bulge}}
\end{figure}

A possible connection between the thick disk
and the Galactic bulge was first presented by \cite{melendez2008} 
who showed that the $\alpha$-element trends of a handful of bulge 
red giants to those of nearby red giants were similar, and hence
that the bulge and the thick disk possibly could have had very
similar chemical and star formation histories. Subsequent studies
of red giants \citep[e.g.,][]{alvesbrito2010,ryde2010,gonzalez2011}
and microlensed dwarf stars \citep{bensby2009,bensby2010,bensby2011,bensby2013} 
confirm the similarities in the abundance trends between the 
(metal-poor) bulge and thick disk. However, the latest paper on 
microlensed
bulge dwarf stars show that there might be a slight
difference in the abundance trends in the sense that the 
position of the ``knee'' in the bulge [$\alpha$/Fe] trend occurs
at a slightly higher [Fe/H], pointing to that the
star formation history was slightly faster in the bulge than
in the thick disk \citep{bensby2013}. Figure~\ref{fig:bulge}
shows the [Fe/Si]-[Si/H] abundance trends for the thin and thick disk
dwarf sample from \cite{bensby2013disk} and the
microlensed bulge dwarfs from \cite{bensby2013}.

Additional signs of a possible connection between the bulge and the 
thick disk comes from that the metallicity distributions are similar
\citep[e.g.,][]{hill2011} and that the ages are old (around or above
10\,Gyr) \citep{bensby2013} for thick disk and the metal-poor bulge.

Whether these connections imply a direct connection between the thick 
disk and bulge, or not, is currently unclear and needs to be further 
investigated.

\section{Summary}

In summary, it appears as if the Milky Way galaxy has two stellar
disk populations that are distinct in abundance space as well as 
in stellar ages. They do, however, appear to show a considerable
overlap in velocity space. It also appears as if the
radial extension of the thick disk is quite short, and that the
thin disk is dominating at larger galactocentric radii.
Based on similarities of elemental abundances and stellar ages
there might also be a connection between the metal-poor 
bulge and the thick disk.
Whether they are directly connected or if they have just experienced 
similar chemical and star formation histories is unclear.

There are several ongoing and upcoming large spectroscopic surveys that 
will probe the abundance structure of the Galactic disk that will help us 
to disentangle different stellar components and unravelling the outer disk.
Examples are the 
SDSS SEGUE \citep{yanni2009}, APOGEE \citep{allendeprieto2008},
the Gaia-ESO Survey \citep{gilmore2012}, the GALAH survey 
\citep[e.g.,][]{zucker2012} that will provide abundance data for several
hundreds of thousands of stars. Combined with the astrometric data from
the upcoming Gaia satellite, we should get a much clearer picture of the
structure, formation, and evolution of the Milky Way and its stellar 
populations. 

\begin{acknowledgments}
T.B. was funded by grant No. 621-2009-3911 from The Swedish 
Research Council.
\end{acknowledgments}


\newpage
\begin{discussion}

\discuss{Juntai Shen}{The bulge and thick disk stars share some similarities. Could they be related somehow?}

\discuss{Thomas Bensby}{See discussion in Sect.~\ref{sec:bulge}.}

\discuss{Johannes Andersen}{Two things bother me:  (1) Getting membership possibilities by modelling the velocity distribution as a sum of two Gaussians seems to be a dubious affair, when the actual velocity distribution (e.g., from the GCS) looks nothing like that.  (2) Stellar ages tend to have uncertainties of $\sim$25\,\% in the best cases, so ages quoted without uncertainties tend to make the picture clearer than it actually is.}

\discuss{Thomas Bensby}{I agree that the Gaussian assumption is not 
correct, but we use it only as a starting point to select {\sl candidate} 
thin and thick disk stars that we later observe with high-resolution 
spectrographs. Based on the succeeding analysis of these spectra we get 
ages and elemental abundances that can further help us to sort out 
the different stellar populations.
Furthermore, most of the small scale structure seen in the UV-plane 
consists of young  stellar streams. Most of these are true members of 
the thin disk and our Gaussian selection criteria are wide enough that 
these young streams are, correctly, classified as thin disk.}

\end{discussion}


\begin{thebibliography}{62}
\expandafter\ifx\csname natexlab\endcsname\relax\def\natexlab#1{#1}\fi

\bibitem[{{Abazajian} {et~al.}(2009){Abazajian}, {Adelman-McCarthy},
  {Ag{\"u}eros}, {Allam}, {Allende Prieto}, {An}, {Anderson}, {Anderson},
  {Annis}, {Bahcall}, \& et~al.}]{abazajian2009}
{Abazajian}, K.~N., {Adelman-McCarthy}, J.~K., {Ag{\"u}eros}, M.~A., {et~al.}
  2009, \apjs, 182, 543

\bibitem[{{Adibekyan} {et~al.}(2012){Adibekyan}, {Sousa}, {Santos}, {Delgado
  Mena}, {Gonz{\'a}lez Hern{\'a}ndez}, {Israelian}, {Mayor}, \&
  {Khachatryan}}]{adibekyan2012}
{Adibekyan}, V.~Z., {Sousa}, S.~G., {Santos}, N.~C., {et~al.} 2012, \aap, 545,
  A32

\bibitem[{{Allende Prieto} {et~al.}(2006){Allende Prieto}, {Beers}, {Wilhelm},
  {Newberg}, {Rockosi}, {Yanny}, \& {Lee}}]{allendeprieto2006}
{Allende Prieto}, C., {Beers}, T.~C., {Wilhelm}, R., {et~al.} 2006, \apj, 636,
  804

\bibitem[{{Allende Prieto} {et~al.}(2008){Allende Prieto}, {Majewski},
  {Schiavon}, {Cunha}, {Frinchaboy}, {Holtzman}, {Johnston}, {Shetrone},
  {Skrutskie}, {Smith}, \& {Wilson}}]{allendeprieto2008}
{Allende Prieto}, C., {Majewski}, S.~R., {Schiavon}, R., {et~al.} 2008,
  Astronomische Nachrichten, 329, 1018

\bibitem[{{Alves-Brito} {et~al.}(2010){Alves-Brito}, {Mel{\'e}ndez}, {Asplund},
  {Ram{\'{\i}}rez}, \& {Yong}}]{alvesbrito2010}
{Alves-Brito}, A., {Mel{\'e}ndez}, J., {Asplund}, M., {Ram{\'{\i}}rez}, I., \&
  {Yong}, D. 2010, \aap, 513, A35

\bibitem[{{Bensby}(2004)}]{bensby2004phdt}
{Bensby}, T. 2004, Ph.D.~Thesis, Lund University

\bibitem[{{Bensby} {et~al.}(2011{\natexlab{a}}){Bensby}, {Ad{\'e}n},
  {Mel{\'e}ndez}, {Gould}, {Feltzing}, {Asplund}, {Johnson}, {Lucatello},
  {Yee}, {Ram{\'{\i}}rez}, {Cohen}, {Thompson}, {Bond}, {Gal-Yam}, {Han},
  {Sumi}, {Suzuki}, {Wada}, {Miyake}, {Furusawa}, {Ohmori}, {Saito},
  {Tristram}, \& {Bennett}}]{bensby2011}
{Bensby}, T., {Ad{\'e}n}, D., {Mel{\'e}ndez}, J., {et~al.} 2011{\natexlab{a}},
  \aap, 533, A134

\bibitem[{{Bensby} {et~al.}(2010{\natexlab{a}}){Bensby}, {Alves-Brito}, {Oey},
  {Yong}, \& {Mel{\'e}ndez}}]{bensby2010letter}
{Bensby}, T., {Alves-Brito}, A., {Oey}, M.~S., {Yong}, D., \& {Mel{\'e}ndez},
  J. 2010{\natexlab{a}}, \aap, 516, L13

\bibitem[{{Bensby} {et~al.}(2011{\natexlab{b}}){Bensby}, {Alves-Brito}, {Oey},
  {Yong}, \& {Mel{\'e}ndez}}]{bensby2011letter}
{Bensby}, T., {Alves-Brito}, A., {Oey}, M.~S., {Yong}, D., \& {Mel{\'e}ndez},
  J. 2011{\natexlab{b}}, \apjl, 735, L46

\bibitem[{{Bensby} \& {Feltzing}(2010)}]{bensby2010rio}
{Bensby}, T. \& {Feltzing}, S. 2010, in IAU Symposium, Vol. 265, IAU Symposium,
  ed. {K.~Cunha, M.~Spite, \& B.~Barbuy}, 300--303

\bibitem[{{Bensby} {et~al.}(2010{\natexlab{b}}){Bensby}, {Feltzing}, {Johnson},
  {Gould}, {Ad{\'e}n}, M., {Mel\'endez}, {Gal-Yam}, {Lucatello}, {Sana},
  {Sumi}, {Miyake}, {Suzuki}, {Han}, {Bond}, \& {Udalski}}]{bensby2010}
{Bensby}, T., {Feltzing}, S., {Johnson}, J.~A., {et~al.} 2010{\natexlab{b}},
  \aap, 512, A41

\bibitem[{{Bensby} {et~al.}(2003){Bensby}, {Feltzing}, \& {Lundstr{\"
  o}m}}]{bensby2003}
{Bensby}, T., {Feltzing}, S., \& {Lundstr{\" o}m}, I. 2003, \aap, 410, 527

\bibitem[{{Bensby} {et~al.}(2004{\natexlab{a}}){Bensby}, {Feltzing}, \&
  {Lundstr{\" o}m}}]{bensby_amr}
{Bensby}, T., {Feltzing}, S., \& {Lundstr{\" o}m}, I. 2004{\natexlab{a}}, \aap,
  421, 969

\bibitem[{{Bensby} {et~al.}(2004{\natexlab{b}}){Bensby}, {Feltzing}, \&
  {Lundstr{\" o}m}}]{bensby2004}
{Bensby}, T., {Feltzing}, S., \& {Lundstr{\" o}m}, I. 2004{\natexlab{b}}, \aap,
  415, 155

\bibitem[{{Bensby} {et~al.}(2005){Bensby}, {Feltzing}, {Lundstr{\" o}m}, \&
  {Ilyin}}]{bensby2005}
{Bensby}, T., {Feltzing}, S., {Lundstr{\" o}m}, I., \& {Ilyin}, I. 2005, \aap,
  433, 185

\bibitem[{{Bensby} {et~al.}(2013{\natexlab{a}}){Bensby}, {Feltzing}, \&
  {Oey}}]{bensby2013disk}
{Bensby}, T., {Feltzing}, S., \& {Oey}, M.~S. 2013{\natexlab{a}}, \aap,
  submitted

\bibitem[{{Bensby} {et~al.}(2009){Bensby}, {Johnson}, {Cohen}, {Feltzing},
  {Udalski}, {Gould}, {Huang}, {Thompson}, {Simmerer}, \&
  {Ad{\'e}n}}]{bensby2009}
{Bensby}, T., {Johnson}, J.~A., {Cohen}, J., {et~al.} 2009, \aap, 499, 737

\bibitem[{{Bensby} {et~al.}(2007{\natexlab{a}}){Bensby}, {Oey}, {Feltzing}, \&
  {Gustafsson}}]{bensby2007letter}
{Bensby}, T., {Oey}, M.~S., {Feltzing}, S., \& {Gustafsson}, B.
  2007{\natexlab{a}}, \apjl, 655, L89

\bibitem[{{Bensby} {et~al.}(2013{\natexlab{b}}){Bensby}, {Yee}, {Feltzing},
  {Johnson}, {Gould}, {Cohen}, {Asplund}, {Mel{\'e}ndez}, {Lucatello}, {Han},
  {Thompson}, {Gal-Yam}, {Udalski}, {Bennett}, {Bond}, {Kohei}, {Sumi},
  {Suzuki}, {Suzuki}, {Takino}, {Tristram}, {Yamai}, \&
  {Yonehara}}]{bensby2013}
{Bensby}, T., {Yee}, J.~C., {Feltzing}, S., {et~al.} 2013{\natexlab{b}}, \aap,
  549, A147

\bibitem[{{Bensby} {et~al.}(2007{\natexlab{b}}){Bensby}, {Zenn}, {Oey}, \&
  {Feltzing}}]{bensby2007letter2}
{Bensby}, T., {Zenn}, A.~R., {Oey}, M.~S., \& {Feltzing}, S.
  2007{\natexlab{b}}, \apjl, 663, L13

\bibitem[{{Bovy} {et~al.}(2012){Bovy}, {Rix}, \& {Hogg}}]{bovy2012}
{Bovy}, J., {Rix}, H.-W., \& {Hogg}, D.~W. 2012, \apj, 751, 131

\bibitem[{{Burstein}(1979)}]{burstein1979}
{Burstein}, D. 1979, \apj, 234, 829

\bibitem[{{Carollo} {et~al.}(2010){Carollo}, {Beers}, {Chiba}, {Norris},
  {Freeman}, {Lee}, {Ivezi{\'c}}, {Rockosi}, \& {Yanny}}]{carollo2010}
{Carollo}, D., {Beers}, T.~C., {Chiba}, M., {et~al.} 2010, \apj, 712, 692

\bibitem[{{Casagrande} {et~al.}(2011){Casagrande}, {Sch{\"o}nrich}, {Asplund},
  {Cassisi}, {Ramirez}, {Melendez}, {Bensby}, \& {Feltzing}}]{casagrande2011}
{Casagrande}, L., {Sch{\"o}nrich}, R., {Asplund}, M., {et~al.} 2011, \aap, 530,
  A138

\bibitem[{{Chen} {et~al.}(2000){Chen}, {Nissen}, {Zhao}, {Zhang}, \&
  {Benoni}}]{chen2000}
{Chen}, Y.~Q., {Nissen}, P.~E., {Zhao}, G., {Zhang}, H.~W., \& {Benoni}, T.
  2000, \aaps, 141, 491

\bibitem[{{Cheng} {et~al.}(2012{\natexlab{a}}){Cheng}, {Rockosi}, {Morrison},
  {Lee}, {Beers}, {Bizyaev}, {Harding}, {Malanushenko}, {Malanushenko},
  {Oravetz}, {Pan}, {Schlesinger}, {Schneider}, {Simmons}, \&
  {Weaver}}]{cheng2012_2}
{Cheng}, J.~Y., {Rockosi}, C.~M., {Morrison}, H.~L., {et~al.}
  2012{\natexlab{a}}, \apj, 752, 51

\bibitem[{{Cheng} {et~al.}(2012{\natexlab{b}}){Cheng}, {Rockosi}, {Morrison},
  {Sch{\"o}nrich}, {Lee}, {Beers}, {Bizyaev}, {Pan}, \&
  {Schneider}}]{cheng2012_1}
{Cheng}, J.~Y., {Rockosi}, C.~M., {Morrison}, H.~L., {et~al.}
  2012{\natexlab{b}}, \apj, 746, 149

\bibitem[{{Comer{\'o}n} {et~al.}(2011){Comer{\'o}n}, {Elmegreen}, {Knapen},
  {Salo}, {Laurikainen}, {Laine}, {Athanassoula}, {Bosma}, {Sheth}, {Regan},
  {Hinz}, {Gil de Paz}, {Men{\'e}ndez-Delmestre}, {Mizusawa},
  {Mu{\~n}oz-Mateos}, {Seibert}, {Kim}, {Elmegreen}, {Gadotti}, {Ho},
  {Holwerda}, {Lappalainen}, {Schinnerer}, \& {Skibba}}]{comeron2011}
{Comer{\'o}n}, S., {Elmegreen}, B.~G., {Knapen}, J.~H., {et~al.} 2011, \apj,
  741, 28

\bibitem[{{Deason} {et~al.}(2013){Deason}, {Belokurov}, {Evans}, \&
  {Johnston}}]{deason2013}
{Deason}, A.~J., {Belokurov}, V., {Evans}, N.~W., \& {Johnston}, K.~V. 2013,
  \apj, 763, 113

\bibitem[{{Edvardsson} {et~al.}(1993){Edvardsson}, {Andersen}, {Gustafsson},
  {Lambert}, {Nissen}, \& {Tomkin}}]{edvardsson1993}
{Edvardsson}, B., {Andersen}, J., {Gustafsson}, B., {et~al.} 1993, \aap, 275,
  101

\bibitem[{{Famaey} {et~al.}(2005){Famaey}, {Jorissen}, {Luri}, {Mayor}, {Udry},
  {Dejonghe}, \& {Turon}}]{famaey2005}
{Famaey}, B., {Jorissen}, A., {Luri}, X., {et~al.} 2005, \aap, 430, 165

\bibitem[{{Feltzing} \& {Gustafsson}(1998)}]{feltzing1998}
{Feltzing}, S. \& {Gustafsson}, B. 1998, \aaps, 129, 237

\bibitem[{{Fuhrmann}(1998)}]{fuhrmann1998}
{Fuhrmann}, K. 1998, \aap, 338, 161

\bibitem[{{Fuhrmann}(2000)}]{fuhrmann2000unpubl}
{Fuhrmann}, K. 2000, unpublished

\bibitem[{{Fuhrmann}(2004)}]{fuhrmann2004}
{Fuhrmann}, K. 2004, Astronomische Nachrichten, 325, 3

\bibitem[{{Fuhrmann}(2008)}]{fuhrmann2008}
{Fuhrmann}, K. 2008, \mnras, 384, 173

\bibitem[{{Fuhrmann}(2011)}]{fuhrmann2011}
{Fuhrmann}, K. 2011, \mnras, 414, 2893

\bibitem[{{Gilmore} {et~al.}(2012){Gilmore}, {Randich}, {Asplund}, {Binney},
  {Bonifacio}, {Drew}, {Feltzing}, {Ferguson}, {Jeffries}, {Micela},
  {Negueruela}, {Prusti}, {Rix}, {Vallenari}, {Alfaro}, {Allende-Prieto},
  {Babusiaux}, {Bensby}, {Blomme}, {Bragaglia}, {Flaccomio}, {Francois},
  {Irwin}, {Koposov}, {Korn}, {Lanzafame}, {Pancino}, {Paunzen},
  {Recio-Blanco}, {Sacco}, {Smiljanic}, {van Eck}, \& {Walton}}]{gilmore2012}
{Gilmore}, G., {Randich}, S., {Asplund}, M., {et~al.} 2012, The Messenger, 147,
  25

\bibitem[{{Gilmore} \& {Reid}(1983)}]{gilmore1983}
{Gilmore}, G. \& {Reid}, N. 1983, \mnras, 202, 1025

\bibitem[{{Gilmore} {et~al.}(1995){Gilmore}, {Wyse}, \& {Jones}}]{gilmore1995}
{Gilmore}, G., {Wyse}, R.~F.~G., \& {Jones}, J.~B. 1995, \aj, 109, 1095

\bibitem[{{Gilmore} {et~al.}(2002){Gilmore}, {Wyse}, \& {Norris}}]{gilmore2002}
{Gilmore}, G., {Wyse}, R.~F.~G., \& {Norris}, J.~E. 2002, \apjl, 574, L39

\bibitem[{{Gonzalez} {et~al.}(2011){Gonzalez}, {Rejkuba}, {Zoccali}, {Hill},
  {Battaglia}, {Babusiaux}, {Minniti}, {Barbuy}, {Alves-Brito}, {Renzini},
  {Gomez}, \& {Ortolani}}]{gonzalez2011}
{Gonzalez}, O.~A., {Rejkuba}, M., {Zoccali}, M., {et~al.} 2011, \aap, 530, A54

\bibitem[{{Gratton} {et~al.}(2000){Gratton}, {Carretta}, {Matteucci}, \&
  {Sneden}}]{gratton2000}
{Gratton}, R.~G., {Carretta}, E., {Matteucci}, F., \& {Sneden}, C. 2000, \aap,
  358, 671

\bibitem[{{Haywood}(2006)}]{haywood2006}
{Haywood}, M. 2006, \mnras, 371, 1760

\bibitem[{{Haywood} {et~al.}(2013){Haywood}, {Di Matteo}, {Lehnert}, {Katz}, \&
  {Gomez}}]{haywood2013}
{Haywood}, M., {Di Matteo}, P., {Lehnert}, M., {Katz}, D., \& {Gomez}, A. 2013,
  arXiv:1305.4663 [astro-ph.GA]

\bibitem[{{Hill} {et~al.}(2011){Hill}, {Lecureur}, {G{\'o}mez}, {Zoccali},
  {Schultheis}, {Babusiaux}, {Royer}, {Barbuy}, {Arenou}, {Minniti}, \&
  {Ortolani}}]{hill2011}
{Hill}, V., {Lecureur}, A., {G{\'o}mez}, A., {et~al.} 2011, \aap, 534, A80

\bibitem[{{Mashonkina} \& {Gehren}(2001)}]{mashonkina2001}
{Mashonkina}, L. \& {Gehren}, T. 2001, \aap, 376, 232

\bibitem[{{Mel{\'e}ndez} {et~al.}(2008){Mel{\'e}ndez}, {Asplund},
  {Alves-Brito}, {Cunha}, {Barbuy}, {Bessell}, {Chiappini}, {Freeman},
  {Ram{\'{\i}}rez}, {Smith}, \& {Yong}}]{melendez2008}
{Mel{\'e}ndez}, J., {Asplund}, M., {Alves-Brito}, A., {et~al.} 2008, \aap, 484,
  L21

\bibitem[{{Nissen} {et~al.}(2002){Nissen}, {Primas}, {Asplund}, \&
  {Lambert}}]{nissen2002}
{Nissen}, P.~E., {Primas}, F., {Asplund}, M., \& {Lambert}, D.~L. 2002, \aap,
  390, 235

\bibitem[{{Nordstr{\" o}m} {et~al.}(2004){Nordstr{\" o}m}, {Mayor}, {Andersen},
  {Holmberg}, {Pont}, {J{\o}rgensen}, {Olsen}, {Udry}, \&
  {Mowlavi}}]{nordstrom2004}
{Nordstr{\" o}m}, B., {Mayor}, M., {Andersen}, J., {et~al.} 2004, \aap, 418,
  989

\bibitem[{{Prochaska} {et~al.}(2000){Prochaska}, {Naumov}, {Carney},
  {McWilliam}, \& {Wolfe}}]{prochaska2000}
{Prochaska}, J.~X., {Naumov}, S.~O., {Carney}, B.~W., {McWilliam}, A., \&
  {Wolfe}, A.~M. 2000, \aj, 120, 2513

\bibitem[{{Reddy} {et~al.}(2006){Reddy}, {Lambert}, \& {Allende
  Prieto}}]{reddy2006}
{Reddy}, B.~E., {Lambert}, D.~L., \& {Allende Prieto}, C. 2006, \mnras, 367,
  1329

\bibitem[{{Reddy} {et~al.}(2003){Reddy}, {Tomkin}, {Lambert}, \& {Allende
  Prieto}}]{reddy2003}
{Reddy}, B.~E., {Tomkin}, J., {Lambert}, D.~L., \& {Allende Prieto}, C. 2003,
  \mnras, 340, 304

\bibitem[{{Ryde} {et~al.}(2010){Ryde}, {Gustafsson}, {Edvardsson},
  {Mel{\'e}ndez}, {Alves-Brito}, {Asplund}, {Barbuy}, {Hill}, {K{\"a}ufl},
  {Minniti}, {Ortolani}, {Renzini}, \& {Zoccali}}]{ryde2010}
{Ryde}, N., {Gustafsson}, B., {Edvardsson}, B., {et~al.} 2010, \aap, 509, A20

\bibitem[{{Schuster} {et~al.}(2006){Schuster}, {Moitinho}, {M{\'a}rquez},
  {Parrao}, \& {Covarrubias}}]{schuster2006}
{Schuster}, W.~J., {Moitinho}, A., {M{\'a}rquez}, A., {Parrao}, L., \&
  {Covarrubias}, E. 2006, \aap, 445, 939

\bibitem[{{Soubiran} {et~al.}(2003){Soubiran}, {Bienaym{\' e}}, \&
  {Siebert}}]{soubiran2003}
{Soubiran}, C., {Bienaym{\' e}}, O., \& {Siebert}, A. 2003, \aap, 398, 141

\bibitem[{{Tautvai{\v s}ien{\.e}} {et~al.}(2001){Tautvai{\v s}ien{\.e}},
  {Edvardsson}, {Tuominen}, \& {Ilyin}}]{tautvaisiene2001}
{Tautvai{\v s}ien{\.e}}, G., {Edvardsson}, B., {Tuominen}, I., \& {Ilyin}, I.
  2001, \aap, 380, 578

\bibitem[{{Tsikoudi}(1979)}]{tsikoudi1979}
{Tsikoudi}, V. 1979, \apj, 234, 842

\bibitem[{{Wyse} {et~al.}(2006){Wyse}, {Gilmore}, {Norris}, {Wilkinson},
  {Kleyna}, {Koch}, {Evans}, \& {Grebel}}]{wyse2006}
{Wyse}, R.~F.~G., {Gilmore}, G., {Norris}, J.~E., {et~al.} 2006, \apjl, 639,
  L13

\bibitem[{{Yanny} {et~al.}(2009){Yanny}, {Rockosi}, {Newberg}, {Knapp},
  {Adelman-McCarthy}, {Alcorn}, {Allam}, {Allende Prieto}, {An}, {Anderson},
  {Anderson}, {Bailer-Jones}, {Bastian}, {Beers}, {Bell}, {Belokurov},
  {Bizyaev}, {Blythe}, {Bochanski}, {Boroski}, {Brinchmann}, {Brinkmann},
  {Brewington}, {Carey}, {Cudworth}, {Evans}, {Evans}, {Gates}, {G{\"a}nsicke},
  {Gillespie}, {Gilmore}, {Nebot Gomez-Moran}, {Grebel}, {Greenwell}, {Gunn},
  {Jordan}, {Jordan}, {Harding}, {Harris}, {Hendry}, {Holder}, {Ivans},
  {Ivezi{\v c}}, {Jester}, {Johnson}, {Kent}, {Kleinman}, {Kniazev},
  {Krzesinski}, {Kron}, {Kuropatkin}, {Lebedeva}, {Lee}, {French Leger},
  {L{\'e}pine}, {Levine}, {Lin}, {Long}, {Loomis}, {Lupton}, {Malanushenko},
  {Malanushenko}, {Margon}, {Martinez-Delgado}, {McGehee}, {Monet}, {Morrison},
  {Munn}, {Neilsen}, {Nitta}, {Norris}, {Oravetz}, {Owen}, {Padmanabhan},
  {Pan}, {Peterson}, {Pier}, {Platson}, {Re Fiorentin}, {Richards}, {Rix},
  {Schlegel}, {Schneider}, {Schreiber}, {Schwope}, {Sibley}, {Simmons},
  {Snedden}, {Allyn Smith}, {Stark}, {Stauffer}, {Steinmetz}, {Stoughton},
  {SubbaRao}, {Szalay}, {Szkody}, {Thakar}, {Sivarani}, {Tucker}, {Uomoto},
  {Vanden Berk}, {Vidrih}, {Wadadekar}, {Watters}, {Wilhelm}, {Wyse}, {Yarger},
  \& {Zucker}}]{yanni2009}
{Yanny}, B., {Rockosi}, C., {Newberg}, H.~J., {et~al.} 2009, \aj, 137, 4377

\bibitem[{{Yoachim} \& {Dalcanton}(2006)}]{yoachim2006}
{Yoachim}, P. \& {Dalcanton}, J.~J. 2006, \aj, 131, 226

\bibitem[{{Zucker} {et~al.}(2012){Zucker}, {de Silva}, {Freeman},
  {Bland-Hawthorn}, \& {Hermes Team}}]{zucker2012}
{Zucker}, D.~B., {de Silva}, G., {Freeman}, K., {Bland-Hawthorn}, J., \&
  {Hermes Team}. 2012, in Astronomical Society of the Pacific Conference
  Series, Vol. 458, Galactic Archaeology: Near-Field Cosmology and the
  Formation of the Milky Way, ed. W.~{Aoki}, M.~{Ishigaki}, T.~{Suda},
  T.~{Tsujimoto}, \& N.~{Arimoto}, 421

\end{thebibliography}
\end{document}